\documentclass[conference]{IEEEtran}
\IEEEoverridecommandlockouts

\usepackage{cite}
\usepackage{amsmath,amssymb,amsfonts}
\usepackage{algorithmic}
\usepackage{graphicx}
\usepackage[dvipsnames]{xcolor}
\usepackage[final]{microtype}
\usepackage[italic]{mathastext}
\usepackage{libertine}
\usepackage[T1]{fontenc}
\usepackage{textcomp}
\usepackage[varqu,varl]{zi4}
\usepackage[all]{nowidow}
\usepackage[keeplastbox]{flushend}
\usepackage{fancyhdr}
\usepackage{xcolor}

\usepackage{caption}
\usepackage{enumitem}
\usepackage{subcaption}
\usepackage{multirow}
\usepackage{mathtools}
\usepackage{xspace}
\DeclarePairedDelimiter\ceil{\lceil}{\rceil}
\DeclarePairedDelimiter\floor{\lfloor}{\rfloor}

\def\BibTeX{{\rm B\kern-.05em{\sc i\kern-.025em b}\kern-.08em
    T\kern-.1667em\lower.7ex\hbox{E}\kern-.125emX}}

\newcommand{\mypar}[1]{\noindent \textbf{#1:}}

\newcommand{\eg}{\textit{e.g.}}
\newcommand{\ie}{\textit{i.e.}}
\newcommand{\etal}{\textit{et al.}\xspace}

\newcommand{\ignore}[1]{}

\newcounter{findingscounter}
\newcommand{\finding}[1]{
    \addtocounter{findingscounter}{1}
    \vspace{0.25cm}
    \noindent\fbox{%
        \parbox{0.97\columnwidth}{%
            \textbf{Finding \thefindingscounter:} #1
        }%
    }
}

\RequirePackage[normalem]{ulem}
\RequirePackage{color}\definecolor{RED}{rgb}{1,0,0}\definecolor{BLUE}{rgb}{0,0,1}
\providecommand{\DIFadd}[1]{{\protect\color{black}#1}}
\providecommand{\DIFaddAlt}[1]{{\protect\color{black}#1}}

\providecommand{\DIFedit}[1]{{\protect\color{black}#1}} 

\begin{document}


\title{Taming Performance Variability caused by\\ Client-Side Hardware Configuration}

\author{\IEEEauthorblockN{Georgia Antoniou, Haris Volos, Yiannakis Sazeides\\
University of Cyprus \\
\{ganton12, hvolos01, yanos\}@ucy.ac.cy
}
}

\maketitle

\begin{abstract}
Many online services running in datacenters are implemented using a microservice software architecture characterized by strict latency requirements. Consequently, this popular software paradigm is increasingly used for the performance evaluation of server systems. Due to the scale and complexity of datacenters, the evaluation of server optimization techniques is usually done on a smaller scale using a client-server model. Although the experimental details of the server side are excessively described in most publications, the client side is often ignored. 
This paper identifies the hardware configuration of the client side as an important source of performance variation
that can affect the accuracy and the correctness of the conclusions of a study that analyzes the performance of microservices. 
This is partially 
attributed to 
the strict latency requirements of microservices and the small scale of the experimental environment. 

In this work we present, using a widely used online-service, several examples where the accuracy and the trends of the conclusions differ based on the configuration of the client-side. At the same time we show that the experimental evaluation time can be significantly affected by the hardware configuration of the client. All these provoke the discussion of the right way to configure the experimental environment for assessing the performance of microservices. 
\end{abstract}


\section{Introduction}
Online applications running in today’s datacenters, such as social networks and web search, have moved from a monolithic to a microservice-based architecture. In this architecture, a monolithic application is decomposed into smaller, interconnected services that communicate explicitly with each other over the network through well-defined interfaces. 
These microservices can be independently developed, deployed, and scaled.
However, due to the increased network overhead arising from the need for communication among services, each service must now adhere to stricter Quality-of-Service (QoS) constraints compared to its monolithic counterpart. 
Previous work reports tight QoS constraints for individual services, with 99th percentile latency targets that range from 250us to 500us
~\cite{chou:udpm:hpca:2019,zhan:carb:cal:2016,belay:ixos:tocs:2016,kasture:tailbench:iiswc:2016}. 

Given the rising prevalence of latency-critical applications based on microservices in today's datacenters, the scientific community has increasingly turned to microservices to evaluate the performance of proposals targeting modern datacenter systems. 
This includes widely-used services, such as Memcached~\cite{memcached}, which is typically deployed as a distributed caching service to accelerate user-facing applications~\cite{berk:facebook-kv-workload:sigmetrics:2012}, and new benchmark suites, such as MicroSuite~\cite{sriraman:usuite:iiswc:2018}, DeathStar~\cite{gan:deathstar:asplos:2019}, and TrainTicket~\cite{trainticket}, which implement representative applications based on microservices.

Experimental evaluation utilizing the above frameworks typically entails deploying them on a small test cluster, following a client-server model. The test cluster often has fewer machine nodes compared to larger-scale production clusters, primarily due to complexity, scale, and cost constraints
~\cite{leverich:reconciling:eurosys:2014,asyabi:peafowl:socc:2020,chou:dynsleep:islped:2016,wang:smartharvest:eurosys:2021}.
Under this deployment, the test cluster comprises a set of server-side and client-side machines. 
Server-side machines are usually configured to host a few services rather than the whole application to keep the test cluster size under control while still achieving a representative setup. 
Client-side machines host the workload generator, which (i) generates a representative workload for the application to process, and (ii) accurately measures the end-to-end performance of the system under a target load, such as average response latency and tail latency (\eg, 99th percentile).  

\begin{table}[t]
\begin{center}
\caption{Hardware characterization in previous work.}
\label{tab:client-side-characterization-survey}
\begin{tabular}{|l|c|}
\hline
\textbf{Characterization} & \textbf{Publications} \\
\hline
Client only & 0 \\
Server only & 8 \\
Client and server & 2 \\
None & 10 \\
\hline
Total & 20 \\
\hline
\end{tabular}
\end{center}
\end{table}

We observe that while experimental evaluations typically specify the server-side hardware configuration to ensure reproducibility of results, they often overlook the client-side configuration.
Table~\ref{tab:client-side-characterization-survey} surveys the client- and server-side hardware configuration in recent publications (\DIFadd{from the years} 2021, 2022, and 2023) across various system and architecture conferences, including ISPASS, IISWC and MICRO.
We find that only 10\% of the papers studied specify the client-side hardware configuration.
We attribute this limitation partially to the implicit assumption that the end-to-end response latency is dominated mainly by the server-side execution time. This assumption is rooted in past practices in experimental evaluations that were based on monolithic applications with millisecond-scale response latencies.
However, this assumption no longer holds with microservices having microsecond-scale response latencies where any client-side microsecond-scale overhead can significantly impact the response latency. 
For example, waking up of a client-side processor core from a power sleep state takes from 2us to 200us~\cite{jawad:agilewatts:micro:2022} (depending on the sleep state). 
This overhead can significantly impact the response latency of a microsecond-scale microservice, such as Memcached with an average server-side processing time of 10us \cite{asyabi:peafowl:socc:2020,chou:udpm:hpca:2019}, 
resulting to an end-to-end response latency that can reach up to hundreds of microseconds. 

To analyze the effect of client-side hardware configuration on performance evaluation, we conduct an experimental study
based on representative, 
\DIFaddAlt{microservice-based services and applications with microseconds/few milliseconds response latencies}, including Memcached~\cite{memcached}, 
HDSearch from MicroSuite~\cite{sriraman:usuite:iiswc:2018}\DIFadd{, Social Network from DeathStar~\cite{gan:deathstar:asplos:2019}} \DIFadd{and synthetic workloads}.
Our experimental analysis reveals that client-side microsecond-scale hardware overheads, such as waking up from a power sleep state and dynamic voltage frequency scaling (DVFS~\cite{sazeides:idvfs:micro:2021}), \DIFaddAlt{can impact the accuracy of the end-to-end measurements leading to incorrect conclusions and additionally introducing performance variation.  
}We find that the extent of the impact depends on a combination of (i) workload generator design, (ii) hardware configuration parameters, and (iii) server-side processing latency. 

This behavior has several ramifications. In an academic setup, an analysis without consideration of the client-side hardware configuration can lead to inaccurate or wrong conclusions. 
At the same time, it renders the work unrepeatable as important details are missing from the experimental methodology of the paper. 
Finally, it degrades the validity of comparisons among techniques optimizing similar metrics in similar environments. 
In an industrial environment, performance evaluation is crucial for determining the load a machine can sustain without any QoS violations and guiding resource allocation for data centers~\cite{nikolaou:tco:bookchapter:2022,nikolaou:tco:tsusc:2022}. Ignoring client-side hardware configuration in this context can result in overprovisioning or underprovisioning of resources.

In summary, we make the following key contributions:
\begin{itemize}[leftmargin=*,noitemsep,topsep=0pt]
\item We identify client-side configuration as a key source of performance variation in experimental evaluation.
\item We demonstrate experimentally how and when client-side configuration can influence the accuracy and validity of the conclusions.
\item We analyse the impact of different client-side configurations on the experimental evaluation time.
\item We provide recommendations for how an experimental environment should be configured to \DIFadd{improve representativeness and thus} mitigate 
measurement inaccuracy caused by client-side configuration.
\end{itemize}
\section{Client-caused Performance Variability}
\label{sec:clientcaused_variation}

\begin{figure}[t]
    \centering
    \includegraphics[width=0.75\linewidth]{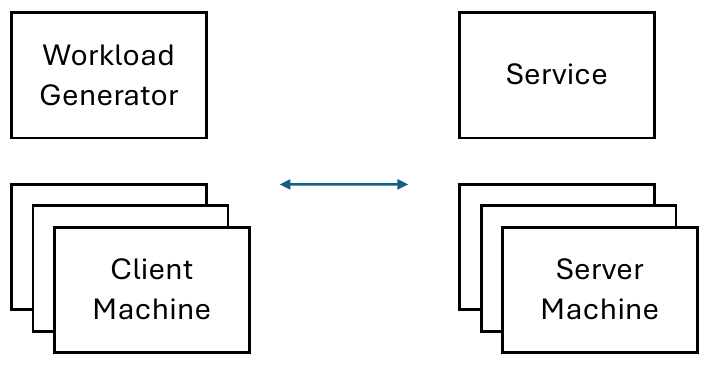}
    \caption{Typical experimental methodology.}
    \label{fig:measurement-methodology}
\end{figure}

Due to the large scale of datacenters and web-based applications, researchers and practitioners typically evaluate data center related optimizations on test clusters with few nodes before propagating the optimization to the rest of the infrastructure.
Measuring the performance of a service typically involves using a workload generator running on a set of client machines, as illustrated in Figure~\ref{fig:measurement-methodology}.

A workload generator is a software component that acts as a client that 
(i) generates requests for the service under study following a representative workload, and 
(ii) accurately measures the end-to-end latency (i.e., average latency, 99th percentile latency). 
Workload generators include the load intensity which represents the inter-arrival time of requests and resource demands which represents characteristics like the type and size of a request.
Most previous work on experimental evaluation focuses on the workload generator (design and configuration), often neglecting \DIFadd{the configuration of} the client machines on which the generator runs.

Our key hypothesis is that client machine configuration can significantly impact workload and measurement accuracy, and the derived conclusions.
The extent of the impact depends on a combination of (i) workload generator design, (ii) hardware configuration parameters, and (iii) service latency. Below, we qualitatively discuss how each such dimension may impact measurement inaccuracy. Later, in Section~\ref{sec:evaluation}, we present empirical evidence supporting our key premise.\\ 

\mypar{Workload generator design} 
A workload generator timestamps generated requests and corresponding replies to model a target workload and measure end-to-end latency. 
This timestamp-based design makes the generator sensitive to timing inaccuracy in two ways.
First, an open-loop generator models an infinite number of requests~\cite{kogias:lancet:usenix:2019}, sending requests to the target service according to an inter-arrival time distribution that represents the time between successive requests. 
Any inaccuracy in timing can disrupt the inter-arrival times, causing requests to shift in time and deviate from the target distribution.
A closed-loop generator further limits the number of outstanding requests to model a finite number of blocking clients~\cite{kogias:lancet:usenix:2019}. 
Because the timing of the next request depends on when the response to the previous request arrives, any timing inaccuracy can further impact the time when a successive request is sent. 
Overall, for both generator types, any timing inaccuracy can impact the timing of requests, causing the generated workload to deviate from the target workload.

Second, a workload generator can measure latency at various points in the system, such as the network interface card (NIC), the in-kernel socket layer, or the generator itself, collectively referred to as points of measurement~\cite{kogias:lancet:usenix:2019}. 
With most typical workload generators, the measurement point resides within the workload generator itself. Therefore, measuring end-to-end latency depends on when the response reaches the generator and when the generator timestamps the response accordingly, rendering the measurement accuracy susceptible to any delay.

\mypar{Hardware configuration parameters} 
The client-side hardware configuration refers to different configuration settings of the client-side system, including hyperthreading, turbo mode, C-states and CPU frequency. 
Such hardware settings can impact timing accuracy, potentially impacting both the generated workload and latency measurements in combination with the design of the workload generator.

For example, consider a time-sensitive workload generator with a point of measurement inside the generator itself, operating on a system with enabled c-states, allowing the system to sleep when idle.
Upon issuing a request, the system may enter a sleep state until the corresponding response arrives. When the response arrives, the system must first wake-up and ramp up its frequency before the workload generator can timestamp the response and measure the end-to-end latency, consequently increasing the measured response time.
Although this scenario may seem straightforward to avoid by configuring the client system to disable hardware features affecting timing accuracy to eliminate any variability, this approach may not always align with the target environment. 
For a target environment enabling c-states for low power, the point of measurement shall include any latency introduced by sleep state transitions. 
Otherwise, the experimental analysis may not be representative in terms of end-to-end latency. Since these types of analysis estimate the speedup of an optimization and ultimately guide the number of resources required to serve a target load, 
an inaccurate experimental environment may cause either overprovision or underprovision of resources.
Unfortunately, enabling c-states with a time-sensitive workload generator to capture a representative point of measurement, may disrupt the generated workload, thus leading to conflicting choices.

\DIFadd{Additionally to HW configuration parameters, kernel parameters (\eg, choice of idle governor), or compiler optimization flags 
can possibly cause similar effects on the accuracy of the end-to-end measurements. In this work we focus only on HW configuration parameters.}

\mypar{Service latency} 
With the emergence of the microservice software paradigm based on which applications are decomposed to several interconnected smaller services, the QoS of latency critical applications like search and social network has reduced significantly from milliseconds to microseconds while the request rates requirements have remained the same. Hardware overheads that are negligible for monolithic applications are now detrimental for the performance of microservice-based applications. As a result, microservices are especially vulnerable to the overhead introduced by the configuration of the client side since it is in the same order as the response time of the microservice (\ie, 
250us). For example C-state transition overhead can take from 2us up to 
\DIFadd{200us} depending on the processor, while legacy DVFS takes several microseconds (\ie, 30us~\cite{sazeides:idvfs:micro:2021}). In the case of monolithic applications or microservice-based applications with higher response time the client side HW configuration shouldn’t affect significantly 
the accuracy.




\section{Statistics Primer}\label{sec:statistics}
In this section we present the background of the statistical methods used in our analysis.




\noindent\textbf{Confidence Intervals (CI):} 
When we display values for summarized datasets such as mean and average it is important to quantify their accuracy. In other words, since we gather empirical statistics in our experiments, confidence intervals (CI)~\cite{le:perfeval:book:2010,kogias:lancet:usenix:2019} offer some confidence that the empirical distribution collected experimentally is close to the actual distribution of the measured population. CI are ranges in which, we are x\% sure that the population mean lies, where x represents the confidence level. A sampled mean of 20, x=95\% and CI of 19.8 - 20.2, means that the true mean of the population distribution lies within 1\% error from the estimated sampled mean. In order to be confident that a mean is higher than another, their CI should not overlap.

Depending on the distribution of the collected samples, we can either use a parametric or a non-parametric CI expression. Parametric expressions assume that the sampled data are derived from a known distribution (i.e., normal/Gaussian) whereas non-parametric expressions assume that the distribution of the sampled data is unknown. Many studies have demonstrated that data collected experimentally in computer systems, do not follow a normal distribution~\cite{kalibera:benchmark:spects:2005, maricq:tamingvariability:usenix:2018,wright:understandingvariation:hpcmpugc:2009}. This is partially inline with what we have observed in our analysis (see Section~\ref{sec:hw_conf_vs_exp_time}). To avoid assumptions of normality we use non-parametric confidence intervals (and other statistical methods) unless noted for the rest of the paper. 

Non-parameric CI are computed based on the median instead of the mean. The following equations are used to compute the confidence intervals bounds for the median. 

\begin{equation}\label{eq:ci_nonparametric1}
    Lower\_bound = \floor{\frac{n - z\sqrt{n}}{2}} 
\end{equation}

\begin{equation}\label{eq:ci_nonparametric2}
    Upper\_bound = \ceil{1 + \frac{n + z\sqrt{n}}{2}}
\end{equation}

Where n is the number of samples in the set and z is the standard score which depends on the target confidence level. 
For a confidence level of 95\%, z equals 1.96. 
Deriving confidence intervals involves first sorting the set of measurements, then using the above equations to determine the indices of the measurements corresponding to the lower and upper bounds of the confidence interval.
The sample’s median should be within the CI bounds.

\noindent\textbf{IID samples:} 
CI require the samples of a set to be independent and identically distributed (iid)~\cite{le:perfeval:book:2010,kogias:lancet:usenix:2019}. In the case of an experiment where the collection metric is latency, the samples are identically distributed since latency measurements come from the same server. Regarding independence, in the analysis presented below we collect one sample per run. In between runs we reset the environment \DIFadd{and} as a result the measured samples are independent. When there is doubt for the iid-ness of samples, several methods can be used with the standard one being autocorrelation. Autocorrelation is a method that calculates the degree of similarity between a time series and a lagged version of itself. The output of the analysis can be anything between -1 and 1, where 1 represents a positive correlation, -1 a negative correlation and values close to 0 indicate no correlation among samples. Other methods used for assessing the iid-ness of samples include Lag-Plots and Turning Point Test.   

\noindent\textbf{Hypothesis Testing - Shapiro-Wilk Test:}    
Hypothesis testing~\cite{liu:hypothesistesting:tkdd:2015} is a systematic procedure used in statistics to assess whether characteristics of a population occur by chance or not. The first step in hypothesis testing, is to define a null hypothesis like for example two populations are equal. Then identify a test statistic that can evaluate the hypothesis. In our analysis we use a Shapiro-Wilk Test~\cite{shapiro-wilk} in order to test whether the sampled data follow a normal distribution. 
Based on the test statistic results a p-value is calculated. P-values represent the probability of finding the observed results of a test statistic if the null hypothesis is true. The p-value is then compared with a significance level, if it is less than the significance level then we reject the null hypothesis. 
Conventionally 5\% and 1\% confidence levels have been used, which means that there is less than 1 in 20 and 1 in 100 chance of being wrong respectively.

\noindent\textbf{Sample Size for Determining Mean/Median:}
The confidence level and accuracy of a CI depends on the number of samples. The higher the number of samples the better the associated confidence level and accuracy. In this section we describe 2 methods (1 parametric~\cite{jain:performanceanalysis:book:1991}, 1 non parametric~\cite{maricq:tamingvariability:usenix:2018}) that can be used to determine what is the minimum required number of samples (repetitions in our case) that are required to achieve a confidence level with a certain accuracy.  

Equation~\ref{eq:iter_parametric}~\cite{jain:performanceanalysis:book:1991} 
calculates the iterations for parametric distributions:

\begin{equation}\label{eq:iter_parametric}
n = (\frac{100zs}{rx})^2
\end{equation}

\noindent where  z is the normal variate of the desired confidence level (1.96 for 95\% confidence), s is the standard deviation, r is the error \% from the mean and x is the mean of the collected samples.

For non-parametric distributions, 
the CONFIRM method~\cite{maricq:tamingvariability:usenix:2018} is used. To calculate the number of repetitions with CONFIRM
: (i) for a set size n, randomly select a subset s <= n and estimate non-parametric CI, (ii) shuffle set, select another subset, and estimate CI, (iii) repeat this procedure c times and then calculate the means for all the lower bounds of CI and upper bounds of CI, and (iv) if error is less or equal to 1\% then size of the subset equals the number of repetitions, otherwise increase subset size and repeat. 
The original CONFIRM paper uses c=200 and s >= 10 assuming that smaller subsets cannot estimate non-parametric CIs reliably.

\section{Experimental Methodology}

\subsection{System} 
To conduct our experiments we use the c220g5 cluster of the Wisconsin site from the CloudLab~\cite{Dmitry:cloudlab:atc:2019} infrastructure. Our baseline system is a 2 socket server with 2 Skylake-based (Intel Xeon Silver 4114) processors. There are 20 physical cores and 40 hardware threads. The nominal frequency is 2.2GHz with the minimum frequency reaching 0.8 GHz and the maximum Turbo Boost frequency 3 GHz. The server is equipped with 192 GB DDR4 DRAM. The operating system used is UBUNTU 18.04.

\subsection{Benchmarks}
Two representative microservice-based latency critical applications are used in the analysis:

\noindent\textbf{Memcached}~\cite{memcached} is a lightweight key-value store that is widely deployed as a distributed caching service to accelerate user-facing applications with strict latency requirements. Memcached has been the focus of numerous studies, including efforts to provide low microsecond-scale tail latency. In our experiments, we run a memcached instance with 10 worker threads pinned on a single socket. We use an extended version of Mutilate~\cite{leverich:reconciling:eurosys:2014}, as a workload generator.
Following the taxonomy of Section~\ref{sec:clientcaused_variation}, Mutilate is an open-loop workload generator; it implements time-sensitive interarrival times using a block-wait event loop that waits for elapsed time, with the point of measurement residing within the generator itself.
We run Mutilate on 5 machines, one for the master process and 4 for the workload-generator clients, establishing a total of 160 connections. We configure the workload generator to recreate the ETC workload from Facebook~\cite{berk:facebook-kv-workload:sigmetrics:2012}.

\noindent\textbf{HDSearch}~\cite{sriraman:usuite:iiswc:2018} is one of the four information-retrieval services of the MicroSuite microservice-based benchmark suite.
HDSearch is an image similarity search service written in C++, which is structured as a three-tier service using RPC for communication between tiers. It returns images from a large dataset whose feature vectors are near to the query’s feature vector. It uses Locality-Sensitive Hash (LSH) tables to traverse the search space of the problem efficiently. We use the accompanying open-loop client, which generates requests with inter-arrival times drawn from a Poisson distribution, as a workload generator.
Following the taxonomy of Section~\ref{sec:clientcaused_variation}, the client is an open-loop workload generator; it implements time-insensitive interarrival times using a busy-wait loop that actively polls for elapsed time, with the point of measurement residing within the generator itself. 
We use 3 machines to run the benchmark, 1 for each type of process: client, midtier and bucket. Our benchmark configuration follows the configuration of the MicroSuite paper~\cite{sriraman:usuite:iiswc:2018}. Finally, we pin the processes onto specific cores to eliminate process migration.

\DIFadd{\noindent\textbf{Social Network} is a microservice-based application from the
DeathStar~\cite{gan:deathstar:asplos:2019} benchmark suite, consisting of multiple interconnected services. 
We deploy the benchmark on a single node using Docker Swarm. We initialize the social graph using the provided small dataset namely "Reed98 Facebook Networks"~\cite{ryan:networkrepo:aaai:2015}. We use the accompanying open-loop client, which is an extended version of the wrk2 workload generator. We configure the client to (i) establish 20 connections with the server, (ii) send requests using an exponential distribution, and (iii) only use read-user-timeline requests. 
Following the taxonomy of Section~\ref{sec:clientcaused_variation}, the client is an open-loop workload generator; it implements time-sensitive interarrival times using a block-wait event loop that waits for elapsed time, with the point of measurement residing within the generator itself. Finally, before each run we fill the database of the application with posts using compose-post queries.     

}
\DIFadd{\noindent\textbf{Synthetic Workload} is a program 
with tunable service latency, implemented to perform a sensitivity analysis. It can accept an 
input parameter, the value of which specifies by how long the processing time of a request should be extended. 
The processing time is implemented using a busy wait loop instead of a normal wait loop to prevent the core from serving other requests, as the additional wait time should be accounted as service time rather than sleep time. 
We run our service instance with 10 worker threads pinned on a single socket. Following the taxonomy of Section~\ref{sec:clientcaused_variation}, the client of the synthetic workload, is an open-loop workload generator; it implements time-sensitive interarrival times using a block-wait event loop that waits for elapsed time, with the point of measurement residing within the generator itself.
}

Unless stated otherwise, each experiment is the average of 50 runs. The duration of each run is 2 minutes. We collect several metrics during the execution of each experiment with the most important one being the average response time and 99th tail latency. We use the non-parametric expressions to calculate CI with a confidence level of 95\%. In each experiment we tune several hardware knobs. The description of each HW knob is mentioned in Section~\ref{sec:hw_knobs} and the scenarios evaluated are mentioned in Section~\ref{sec:client_conf}.  

\subsection{Hardware Knobs}\label{sec:hw_knobs}

In this section we describe the different HW knobs of the analysis and how we tune them.

\noindent\textbf{C-states}~\cite{gough:powermanagement:book:2015} are power saving states that enable a core to reduce it’s power consumption during idle periods. Skylake-based processors support 4 C-states C0, C1, C1E and C6. We use the intel\_idle.max\_cstate flag and the idle=poll flag to enable/disable any C-states through the grub file. 

\noindent\textbf{Frequency Driver}~\cite{cpufreq} is a component of the CPUFreq subsystem of Linux that enables the OS to scale the frequency and voltage. The frequency driver is responsible for communicating the Frequency/Voltage settings to the hardware. Usually a linux system supports 2 frequency drivers, intel\_pstate and acpi-cpufreq. We pass the intel\_pstate flag to the grub file to enable/disable them.

\noindent\textbf{Frequency Governor}~\cite{cpufreq} is also a component of the CPUFreq subsystem. It is the component responsible to decide the suitable frequency/voltage of the system based on some heuristics. We use cpupower, which is a tool that act as a wrapper around the sysfs kernel interface to specify a frequency driver.

\noindent\textbf{Turbo mode}~\cite{gough:powermanagement:book:2015} is a feature in modern processors that \DIFadd{allows} CPU to dynamically increase it’s clock speed above its nominal frequency under certain conditions (i.e., thermal capacity, number of active cores). We use the Model Specific Register (MSR) 0x1a0 to enable/disable turbo mode.

\noindent\textbf{Simultaneous Multithreading (SMT)}~\cite{tullsen:smt:sigarch:1995} is a feature in modern processors that allow multiple threads to execute on the same physical core at the same time. We use the sys interface to enable/disable this feature.

\noindent\textbf{Uncore Frequency}~\cite{gough:powermanagement:book:2015} refers to the operating frequency of the uncore components of the CPU. These components include the Last Level Cache (LLC), IO interfaces etc. We use the MSR 0x620 to tune the uncore frequency.   

\noindent\textbf{Tickless}~\cite{tickless} is a characteristic of kernels that do not omit clock-scheduling interrupts during idle periods. We pass the nohz flag to the grub file to enable/disable this feature. 

\subsection{Client/Server Configuration}\label{sec:client_conf}
In our analysis we use 2 configurations for the client-side, the low-power (LP) configuration and the high-performance (HP) configuration. The LP configuration represents the default configuration of the system and thus the case where a user is agnostic of the client-side configuration. The HP configuration represents a configuration tuned empirically to achieve high performance.  The details of the configuration can be found in Table~\ref{tab:hw_conf}.

The server side baseline configuration is presented also in Table~\ref{tab:hw_conf}. We choose empirically a configuration that does not introduce high variability and achieves good performance. In the experimental evaluation whenever a HW knob of the server side changes it is explicitly mentioned.

\begin{table}[ht]
\caption{Client- and server-side hardware configurations}
\label{tab:hw_conf}
\resizebox{\columnwidth}{!}{%
\begin{tabular}{|c|cc|c|}
\hline
\textbf{}                                                             & \multicolumn{2}{c|}{\textbf{Client Side}}        & \textbf{Server Side} \\ \hline
\textbf{Configuration}                                                & \multicolumn{1}{c|}{\textbf{LP}}  & \textbf{HP}  & \textbf{Baseline}    \\ \hline
\textbf{C-states}                                                     & \multicolumn{1}{c|}{C0,C1,C1E,C6} & off          & C0,C1                \\ \hline
\textbf{\begin{tabular}[c]{@{}c@{}}Frequency\\ Driver\end{tabular}}   & \multicolumn{1}{c|}{intel pstate} & acpi cpufreq & acpi cpufreq         \\ \hline
\textbf{\begin{tabular}[c]{@{}c@{}}Frequency\\ Governor\end{tabular}} & \multicolumn{1}{c|}{powersave}    & performance  & performance          \\ \hline
\textbf{Turbo}                                                        & \multicolumn{1}{c|}{on}           & on           & off                  \\ \hline
\textbf{SMT}                                                          & \multicolumn{1}{c|}{on}           & on           & off                  \\ \hline
\textbf{\begin{tabular}[c]{@{}c@{}}Uncore\\ Frequency\end{tabular}}   & \multicolumn{1}{c|}{dynamic}      & fixed        & fixed                \\ \hline
Tickless                                                              & \multicolumn{1}{c|}{\DIFadd{off}}           & off          & on                   \\ \hline
\end{tabular}%
}
\end{table}

Table~\ref{tab:scenarios_tested} describes the scenarios tested in the experimental analysis (Section~\ref{sec:evaluation}) using the terminology introduced in Section~\ref{sec:clientcaused_variation}. The last column of the table, indicates which one of the scenarios can cause wrong conclusions (\eg \ X) and the sections each scenario is evaluated in.

\begin{table}[ht]
\caption{Scenarios Tested in Section \ref{sec:evaluation}.}
\label{tab:scenarios_tested}
\begin{tabular}{|cc|c|c|c|}
\hline
\multicolumn{2}{|c|}{\textbf{\begin{tabular}[c]{@{}c@{}}Workload Generator\\ Design\end{tabular}}}                                                             & \multirow{2}{*}{\textbf{\begin{tabular}[c]{@{}c@{}}Client \\ Conf.\end{tabular}}} & \multirow{2}{*}{\textbf{\begin{tabular}[c]{@{}c@{}}Response\\ Time\end{tabular}}} & \multirow{2}{*}{\textbf{\begin{tabular}[c]{@{}c@{}}Risk/\\ Section\end{tabular}}} \\ \cline{1-2}
\multicolumn{1}{|c|}{\textbf{\begin{tabular}[c]{@{}c@{}}inter.\\ rate\end{tabular}}}       & \textbf{\begin{tabular}[c]{@{}c@{}}point of\\ meas.\end{tabular}} &                                                                                   &                                                                                   &                                                                                   \\ \hline
\multicolumn{1}{|c|}{\begin{tabular}[c]{@{}c@{}}open-loop\\ time-sensitive\end{tabular}}   & in-app                                                            & tuned                                                                             & small                                                                             & (5.1,5.3)                                                                         \\ \hline
\multicolumn{1}{|c|}{\begin{tabular}[c]{@{}c@{}}open-loop\\ time-sensitive\end{tabular}}   & in-app                                                            & not-tuned                                                                         & small                                                                             & X(5.1,5.3)                                                                        \\ \hline
\multicolumn{1}{|c|}{\begin{tabular}[c]{@{}c@{}}open-loop\\ time-insensitive\end{tabular}} & in-app                                                            & tuned                                                                             & big                                                                               & (5.2)                                                                             \\ \hline
\multicolumn{1}{|c|}{\begin{tabular}[c]{@{}c@{}}open-loop\\ time-insensitive\end{tabular}} & in-app                                                            & not-tuned                                                                         & big                                                                               & (5.2)                                                                             \\ \hline
\end{tabular}
\end{table}
\section{Experimental Analysis}
\label{sec:evaluation}

We study the impact of client-side hardware configuration on performance 
\DIFadd{variation} under different scenarios (Section~\ref{sec:evaluation:client-conf-impact}) and how this impact varies with services with higher response time (Section~\ref{sec:evaluation:service-latency}). 
Finally, we examine the impact of the client-side configuration on the execution time of the experimental evaluation (Section~\ref{sec:hw_conf_vs_exp_time}).   

\subsection{Client-side Configuration Impact}
\label{sec:evaluation:client-conf-impact}

We present two case studies that aim to evaluate the impact of two server-side features, specifically SMT and C-states, on the performance of the Memcached service. 
Our findings demonstrate that the choice of client-side configuration can lead to varying performance results and differing conclusions regarding the effects of the features under study.

\begin{figure*}[ht]
    \centering
    \includegraphics[width=1\linewidth]{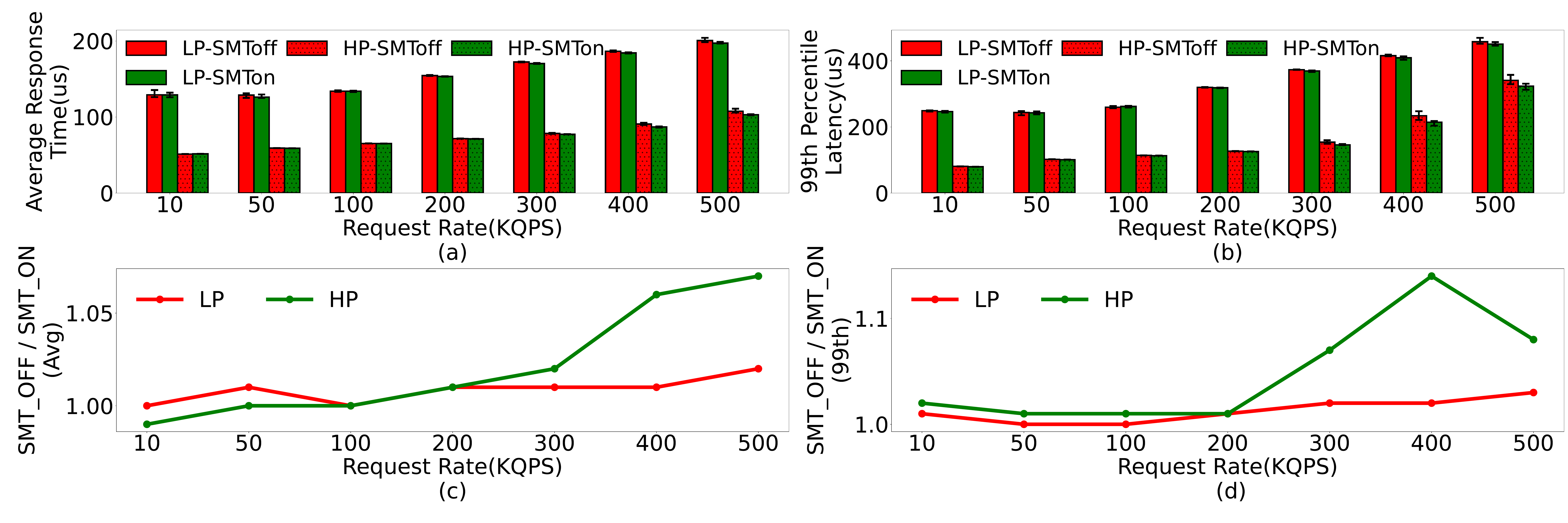}
    \caption{Performance evaluation of SMT impact on Memcached service latency with LP and HP clients. (a) Average Response Time (median) for HP/LP client and SMT ON/OFF server, (b) 99th Percentile Latency (median) for HP/LP client and SMT ON/OFF server, (c) Slowdown (avg) caused by disabling SMT on the Average Response Time for HP and LP client and (d) Slowdown (avg) caused by disabling SMT on the 99th Percentile Latency for HP and LP client.}
    \label{fig:smt:memcached}
\end{figure*}

\mypar{SMT}
The aim of the analysis is to investigate whether SMT can improve the performance of Memcached under different load (10K - 500K QPS) and corresponding utilization (5\% - 55\%). 
Figure~\ref{fig:smt:memcached} shows the performance evaluation of Memcached running on a server machine configured with SMT disabled (baseline) or SMT enabled. Each server-side hardware configuration is examined with two client configurations, namely LP (low power) and HP (high performance).


Depending on the client configuration, the end-to-end measurements differ.
Specifically, the LP end-to-end measurements are between 80\% to 150\% higher than the end-to-end measurements of HP client, that is if we compare similar server-side configurations. Additionally, the  99th percentile latency is 33\% to 200\% higher for LP clients compared to HP clients. 
We argue that this is a result of the additional overhead introduced by the client-side hardware configuration. Since the point of measurement of the workload generator is inside the generator itself, a query must experience at least a C-state transition (2us - 
\DIFadd{200}us), a DVFS transition ($\sim$
\DIFadd{30}us), and a context switch ($\sim$25us) before the workload generator is able to capture the timestamp that will mark the completion of the query. This behavior is especially important in a datacenter setup. Let us assume a service with a QoS of 99th percentile latency equal to 400us. 
The LP client finds that the service can handle only 300K queries without violating any QoS constraints. 
In contrast, the HP client finds that the service can handle 500K queries.
In other words, the LP client determines that a deployment will require 1.6x \DIFaddAlt{more machines than} the HP client, to satisfy the same load without violating any QoS constraints. 

Another important observation from Figure~\ref{fig:smt:memcached}c and Figure~\ref{fig:smt:memcached}d is that the measured degradation depends on the client-side hardware configuration. 
The LP client determines that enabling SMT on the server side improves the 99th percentile latency by at most 3\% (see Figure~\ref{fig:smt:memcached}d). 
In contrast, the HP client determines that the 99th percentile latency can improve by 13\%. We believe this is partially because the absolute performance improvement caused by SMT is more pronounced for the HP client end-to-end time compared to the LP client.


\finding{
    The client-side hardware configuration can impact the accuracy of an experiment. 
    Specifically, it can 
    (i) affect the end-to-end measurements, leading to higher or lower measurements, and 
    (ii) produce different speedups for the same feature or technique under study. 
}

\begin{figure*}[ht]
    \centering
    \includegraphics[width=1\linewidth]{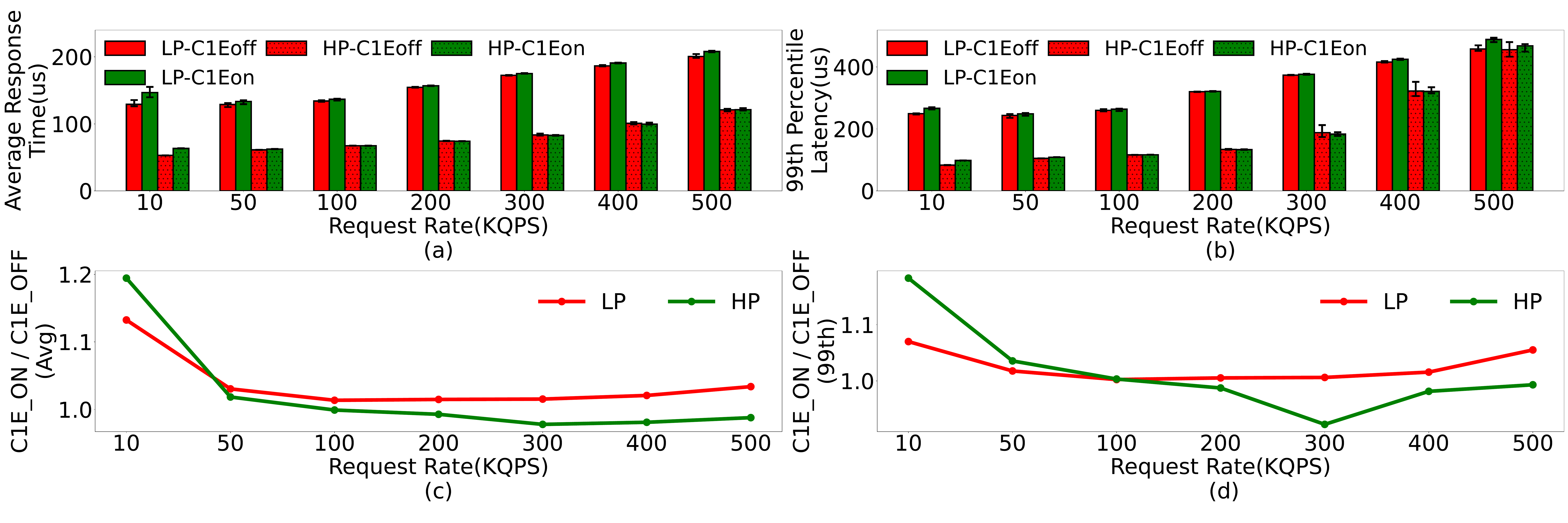}
    \caption{Performance evaluation of C1E impact on Memcached service latency with LP and HP clients. (a) Average Response Time (median) for HP/LP client and C1E ON/OFF server, (b) 99th Percentile Latency (median) for HP/LP client and C1E ON/OFF server, (c) Slowdown (avg) caused by enabling C1E on the Average Response Time for HP and LP client and (d) Slowdown (avg) caused by enabling C1E on the 99th Percentile Latency for HP and LP client.}
    \label{fig:cstates:memcached}
\end{figure*}

\mypar{C1E}
The aim of the analysis is to investigate whether C1E can improve the performance of Memcached under different load (10K - 500K QPS) and corresponding utilization (5\% - 55\%). 
Figure~\ref{fig:cstates:memcached} shows the performance evaluation of Memcached running on a server machine configured with C1E disabled (baseline) or C1E enabled. Each server-side hardware configuration is examined with two client configurations,  namely LP (low power) and HP (high performance).

Similarly to the SMT study above, the choice of client configuration leads to different end-to-end average response latency and 99th percentile latency. Specifically, the average response latency differs from 64\% to 145\% and the 99th percentile latency from 0\% to 200\%. Additionally, the observed slowdown caused by C1E differs based on the client configuration. For the HP client, the slowdown of C1E goes up to 19\% for average latency and 18\% for the 99th percentile latency. For the LP client, the slowdown caused by C1E goes up to 13\% for the average latency and 7\% for the 99th percentile latency. 

More importantly, the client choice shows different trends for high load (400K and 500K QPS), leading to conflicting conclusions about the effect of C1E on performance. The LP client reports that for high load the C1E enabled configuration is worse than the C1E disabled (since the confidence intervals do not overlap). However, the HP client reports that for all loads (except of the 10K QPS load) the C1E enabled and C1E disabled configurations have the same performance.   


\finding{
The client-side hardware configuration can impact not only the accuracy but also the observed trends of an experiment, leading to conflicting conclusions. 
}

\subsection{Impact Relative to Service Latency}
\label{sec:evaluation:service-latency}

\DIFadd{We examine the impact of client-side hardware configuration on the performance of applications with different end-to-end latencies. We present three studies: (i) a single-service study, which investigates the performance of a microservice-based service benchmark, (ii) a multi-service application study, which investigates the performance of a microservice-based application consisting of multiple services, and (iii) a synthetic workload study, which performs a sensitivity analysis. Our findings demonstrate that the client-side hardware configuration has minimal impact on services with high response latency.
}

\begin{figure*}[ht]
    \centering
    \includegraphics[width=1\linewidth]{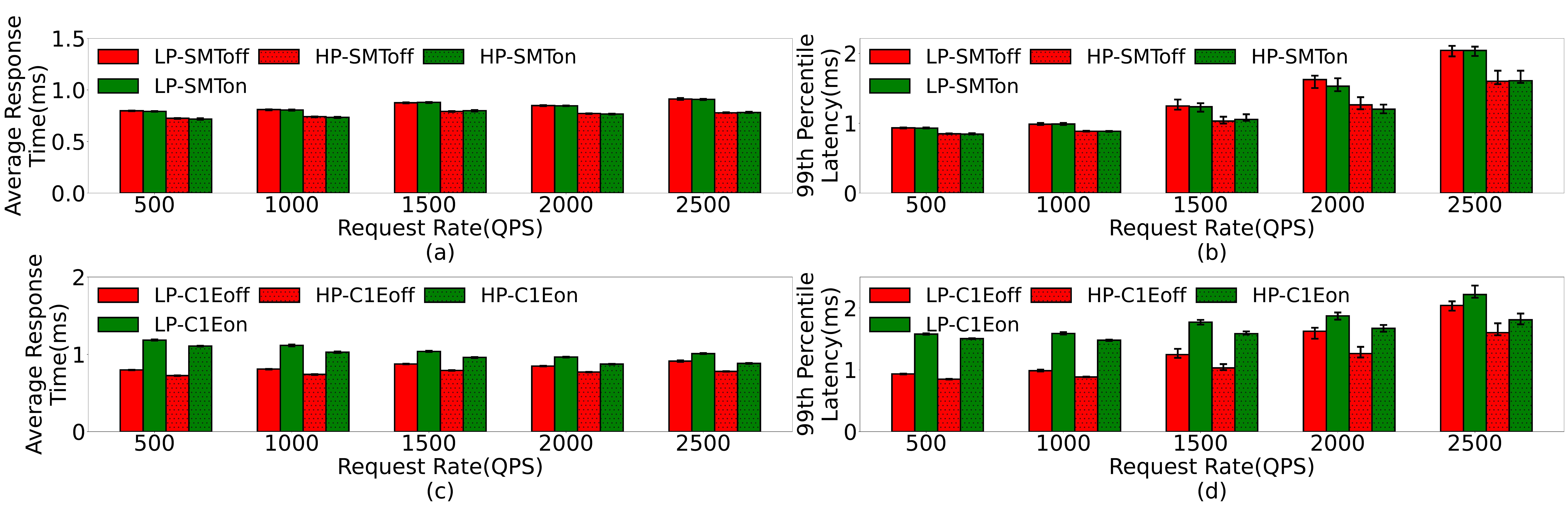}
    \caption{Performance evaluation of SMT and C1E impact on HDSearch service latency with LP and HP clients. (a) Average Response Time (median) for HP/LP client and SMT ON/OFF server, (b) 99th Percentile Latency (median) for HP/LP client and SMT ON/OFF server, (c) Average Response Time (median) for HP/LP client and C1E ON/OFF server and (d) 99th Percentile Latency (median) for HP/LP client and C1E ON/OFF server.}
    \label{fig:smt-cstates:hdsearch}
\end{figure*}

\DIFadd{\mypar{Single-Service}} We \DIFadd{use} 
the HDSearch service, which operates with millisecond-scale latency, to examine the impact of client-side configuration on performance 
\DIFadd{variation} when analyzing the performance of services with high response latency.
Figure~\ref{fig:smt-cstates:hdsearch}, presents the performance evaluation of HDSearch running on a server machine configured with SMT or C1E.
Each server-side HW configuration is examined with two client configurations, namely LP (low power) and HP (high performance).

Similarly to Memcached, there is a difference in end-to-end measurements between the HP and LP client for both average and 99th percentile latency, although it is not as pronounced as in Memcached. Specifically, the average response latency of LP is from 7\% to 17\% higher than HP. Regarding the 99th percentile latency, LP has 5\% to 29\% higher 99th percentile latency than HP. Since HDSearch has higher response latency than Memcached, we expect the difference between the end to end measurements of the HP and LP clients which is a result of the client configuration overhead, to be statistically less significant. 
Thus, we expect LP and HP clients determine similar resource requirements to satisfy a target load without violating any QoS constraints.  


\begin{figure}[ht]
    \centering
    \includegraphics[width=1\linewidth]{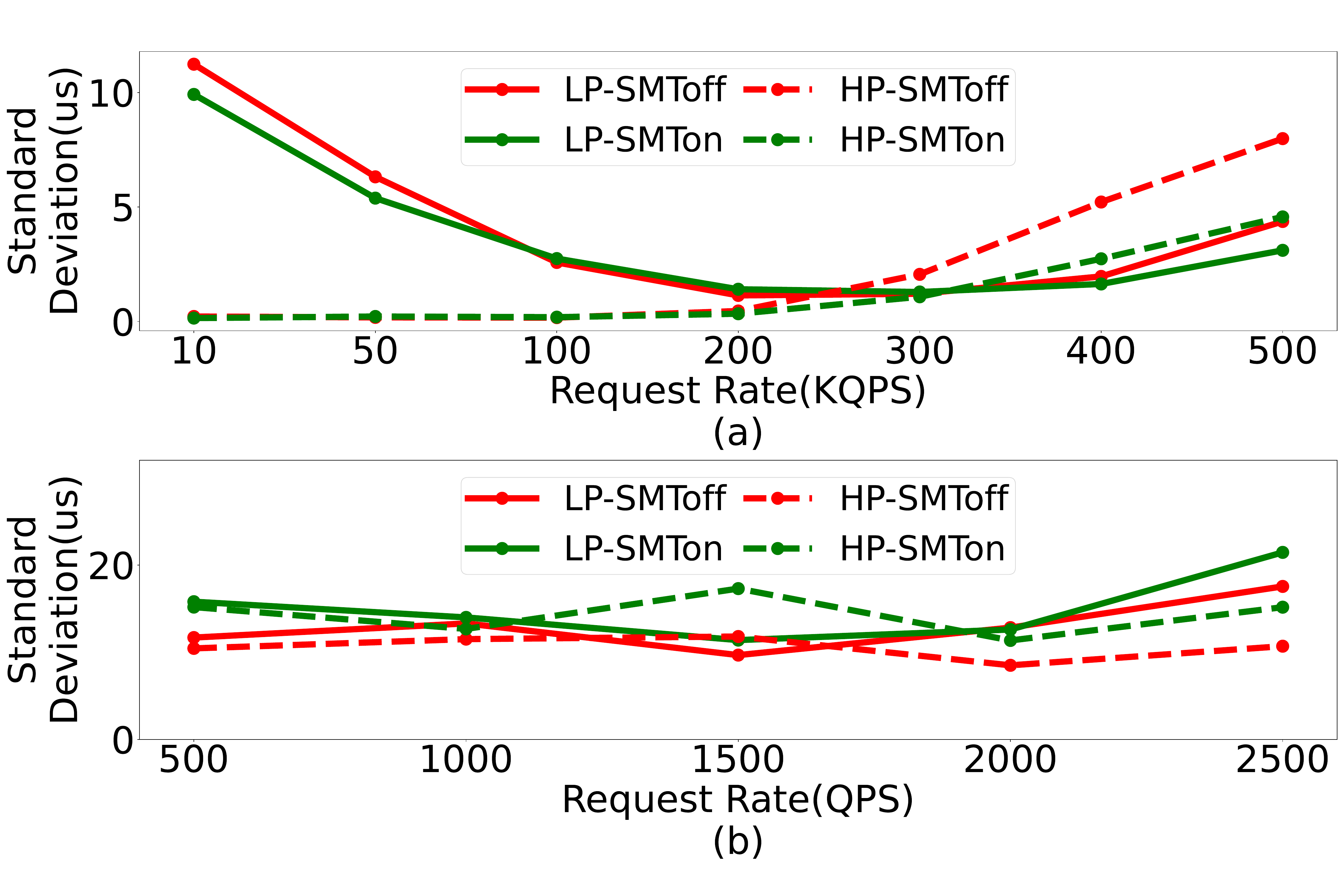}
    \caption{(a) Standard Deviation of Memcached for the Average Response Time with LP/HP client configuration and SMT ON/OFF server configuration, (b) Standard Deviation of HDSearch for the Average Response Time with LP/HP client configuration and SMT ON/OFF server configuration.}
    \label{fig:smt-stdev-avg:mem-hds}
\end{figure}

Contrary to Memcached, the HP and LP clients measure same speedups (with similar trends) in the average response latency for both the SMT and C1E server-side configurations.
Even though the LP measurements experience variability because of the client-side hardware configuration, 
the high server-side processing time of HDSearch (400us) overshadows the client-caused variability ($\sim$20us in  Figure~\ref{fig:smt-stdev-avg:mem-hds}b), thus minimally impacting the observed speedup of the evaluated server configurations. 
Memcached server-side processing time ($\sim$10us) is in the same order as the client-caused variability (up to 10us in Figure~\ref{fig:smt-stdev-avg:mem-hds}a), thus making the speedup of the evaluated server configurations 
more sensitive to the client-side hardware configuration. 



Overall, we observe that HDSearch, a service with about 10 times higher end-to-end response latency than Memcached exhibits similar speedups and trends when run with two different client configurations. Although the absolute end-to-end measurements are not the same, the difference is not as pronounced as in Memcached. 

\begin{figure}[ht]
    \centering
    \includegraphics[width=1\linewidth]{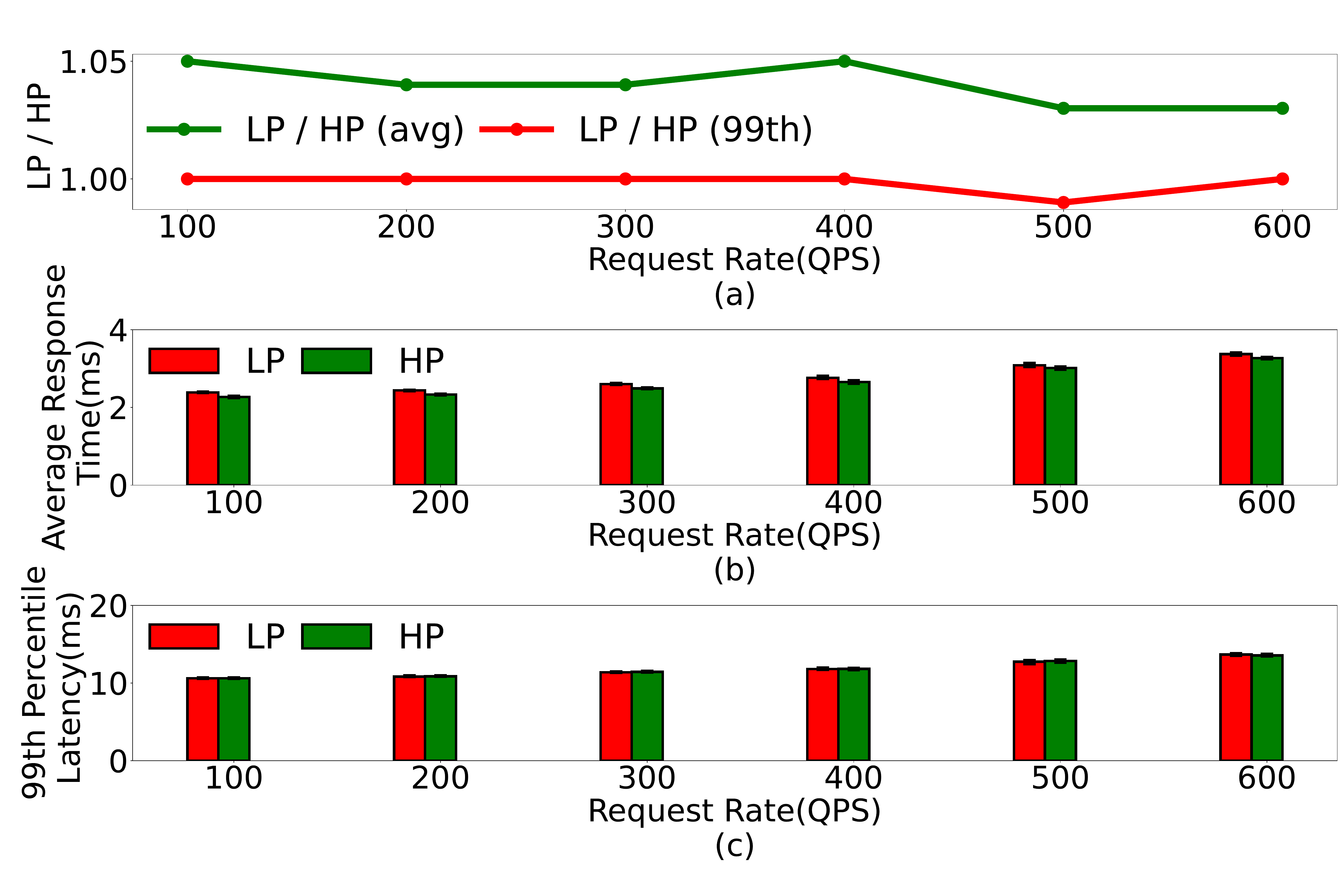}
    \caption{\DIFadd{Performance evaluation of HP and LP clients for Social Network. (a) Slowdown (avg) caused by changing from HP to LP client on the Average Response Time and 99th Percentile Latency, (b) Average Response Time Latency (median) for HP/LP client at different QPS and (c) 99th Percentile Latency (median) for HP/LP client at different QPS. }}
    \label{fig:deathstar:hp_lp}
\end{figure}

\DIFadd{\mypar{Multi-Service Application}} \DIFadd{We study the impact of the client-side HW configuration on the performance of an 
application using Social Network from the DeathStar benchmark suite. 
Figure~\ref{fig:deathstar:hp_lp}a  presents the difference in the end-to-end latency between the two client configurations LP and HP for average and 99th percentile latency respectively.  Similarly to HDSearch the difference between the two clients gets smaller while the end-to-end latency increases. Compared to HDSearch, the gap between the two clients is small (5\% vs 17\% ) on the average response time due to the fact that Social Network has higher end-to-end latency ($\sim$2-3ms in Figure~\ref{fig:deathstar:hp_lp}b) than HDSearch ($\sim$1ms in Figure~\ref{fig:smt-cstates:hdsearch}). Surprisingly the impact of different clients on the 99th percentile latency for Social Network, as shown in Figure~\ref{fig:deathstar:hp_lp}c, is minimal. In other words, the 99th percentile reported by both clients, LP and HP is the same. For end-to-end latencies over 10ms, the client-induced overhead does not appear to affect the accuracy of the measurements, as illustrated in Figure~\ref{fig:deathstar:hp_lp}c.  

Overall, we observe that DeathStar validates the HDSearch analysis. Although the absolute measurements reported by the two clients are not identical, the difference is not as pronounced as in Memcached and HDSearch.}

\DIFadd{\mypar{Synthetic Workload}} \DIFadd{To examine the impact of client side HW configuration at different latencies, we use the synthetic workload. 
Figure~\ref{fig:synthetic:hp_lp}a and Figure~\ref{fig:synthetic:hp_lp}b present the performance evaluation of the synthetic workload 
for different end-to-end latencies and QPS under two client configurations LP and HP. Although the QPS presented in Figure~\ref{fig:synthetic:hp_lp} are low compared to the ones examined in previous sections, it is important to note that due to the increase in processing time there is no opportunity to achieve higher throughput. To determine the examined QPS we use Little’s law and examine only the QPS where the concurrency is less than the number of available cores (i.e., 10) for all possible values of the new parameter. Additionally, the results presented in this section are the average of 20 runs.

As expected with the increase of the end-to-end latency, the gap between the LP and HP reported end-to-end measurements gets smaller. Specifically, the difference goes from 2.8x for 0 added delay to 1.02x for 400 us added delay at 20K QPS. Similarly, for  99th percentile latency the difference goes from 3.5x to 1x. Between 0 to 100us added delay we observe the highest decrease in the difference between the end to end measurements reported by HP and LP client for both average and 99th percentile. This analysis confirms the findings of HDSearch, based on which the added client side overhead becomes less statistically significant for benchmarks with higher latencies.
 
\begin{figure*}[ht]
    \centering
    \includegraphics[width=1\linewidth]{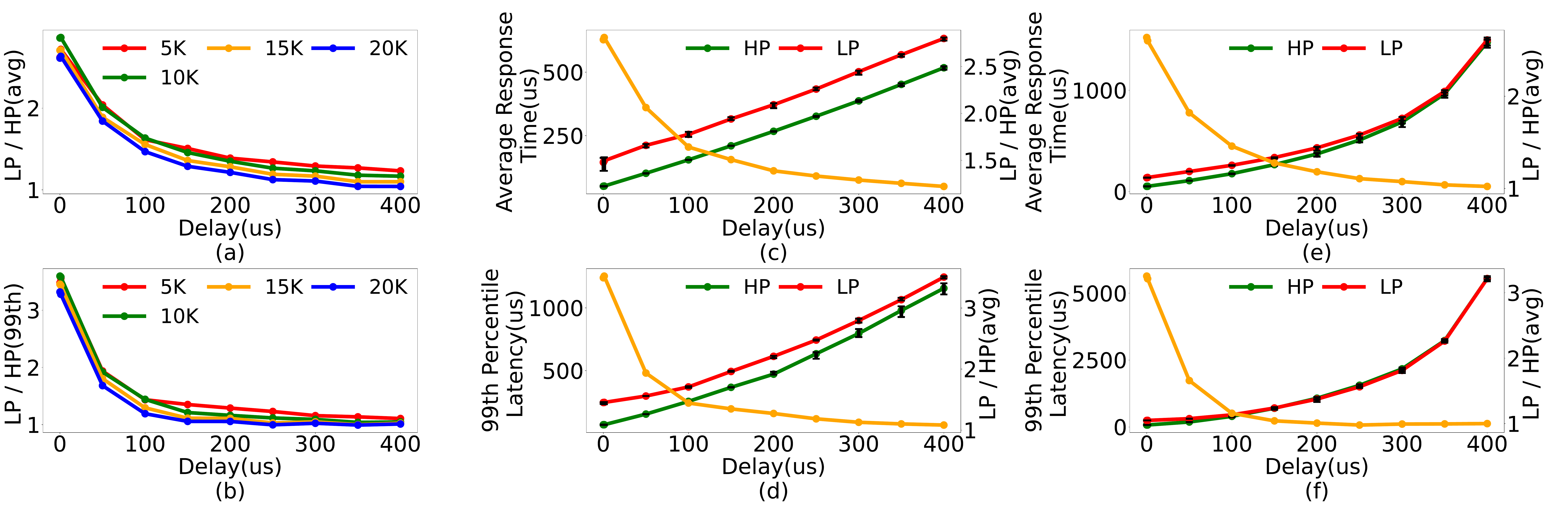}
    \caption{\DIFadd{Performance evaluation of HP and LP clients for different processing times. (a) Slowdown (avg) caused by changing from HP to LP client on the Average Response Time, (b) Slowdown (avg) caused by changing from HP to LP client on the 99th Percentile Latency, (c) Average Response Time (median) for HP/LP client at 5K QPS, (d) 99th Percentile Latency (median) for HP/LP client at 5K QPS, (e) Average Response Time (median) for HP/LP client at 20K QPS and (f) 99th Percentile Latency (median) for HP/LP client at 20K QPS.}}
    \label{fig:synthetic:hp_lp}
\end{figure*}

Figure~\ref{fig:synthetic:hp_lp}c, ~\ref{fig:synthetic:hp_lp}d, ~\ref{fig:synthetic:hp_lp}e and ~\ref{fig:synthetic:hp_lp}f demonstrate the absolute end-to-end measurements for HP and LP client at 5K and 20K QPS.  At low QPS, where there is no queueing, the response time increases linearly with the increase of the added delay which validates the implementation of the synthetic workload. We observe that for average response time latencies over 1ms the accuracy difference is less than 10\% between HP and LP client. 
For high QPS and high added delays (end to end over 2ms) the HP and LP clients measurements converge. This is partially because of the variability of the experiment being in the same order as the observed overhead introduced by the client side hardware configuration (stdev$\sim$100us). A major source of variability for high QPS is the queueing caused in the server side. We conclude that when the end to end latency is in the order of milliseconds, the impact of client side overhead is less significant. 
}

\finding{
    The client-side hardware configuration has minimal impact on services with high response latency.
    The client-side hardware configuration causes performance variability when the processing time of an application is in the same order of magnitude as the variability introduced by the client side.
}

\subsection{Impact on Experimental Evaluation Time}\label{sec:hw_conf_vs_exp_time}

In this section, we investigate how the different client-side hardware configurations affect the experimental evaluation time. By experimental evaluation time, we mean the time required for an experiment to achieve a confidence interval with at most 1\% error at a 95\% confidence level. Before estimating the number of repetitions an experiment requires to gain statistical confidence, we first check whether the collected samples follow a normal distribution. This is because the closed-form expressions used to calculate the number of iterations for an experiment assume that data follow a normal distribution. 

\begin{figure}[ht]
    \centering
    \includegraphics[width=1\linewidth]{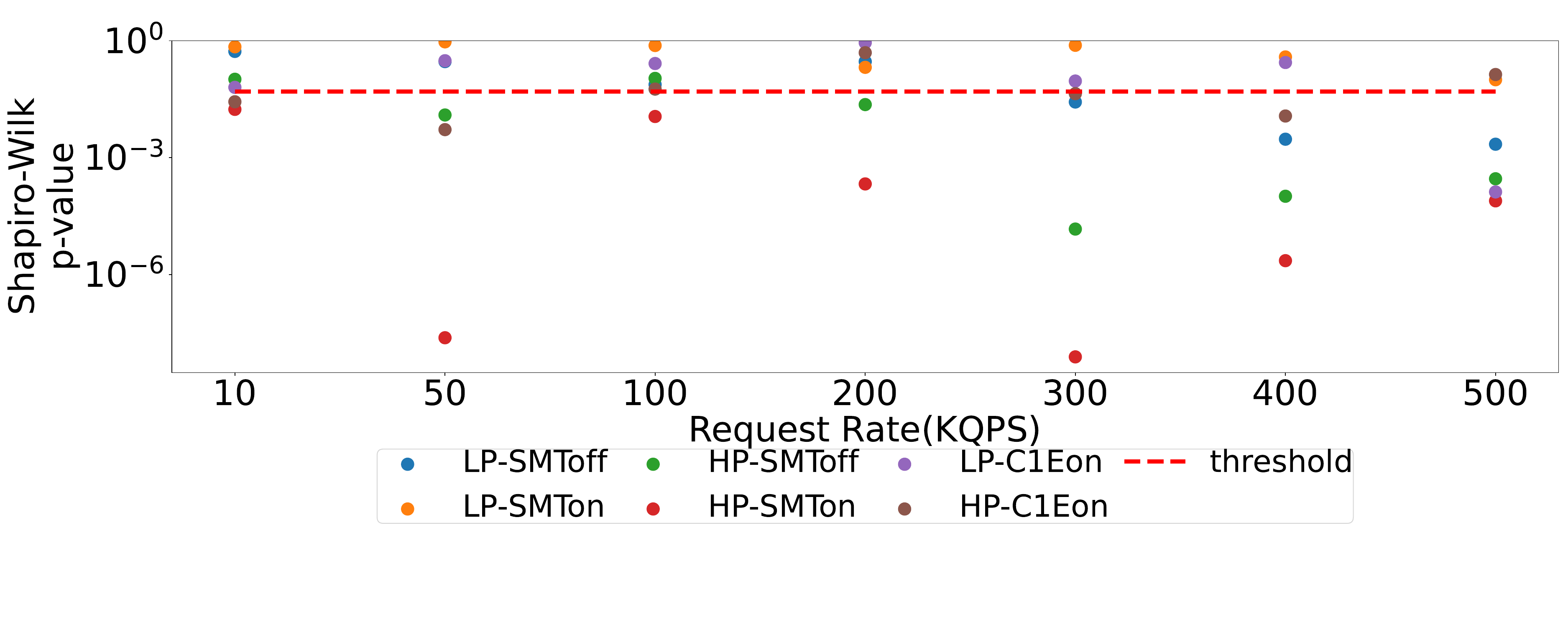}
    \caption{Shapiro-Wilk p-value for configurations in Section~\ref{sec:evaluation:client-conf-impact}.}
    \label{fig:shapiro}
\end{figure}

\begin{figure}[ht]
    \centering
    \includegraphics[width=1\linewidth]{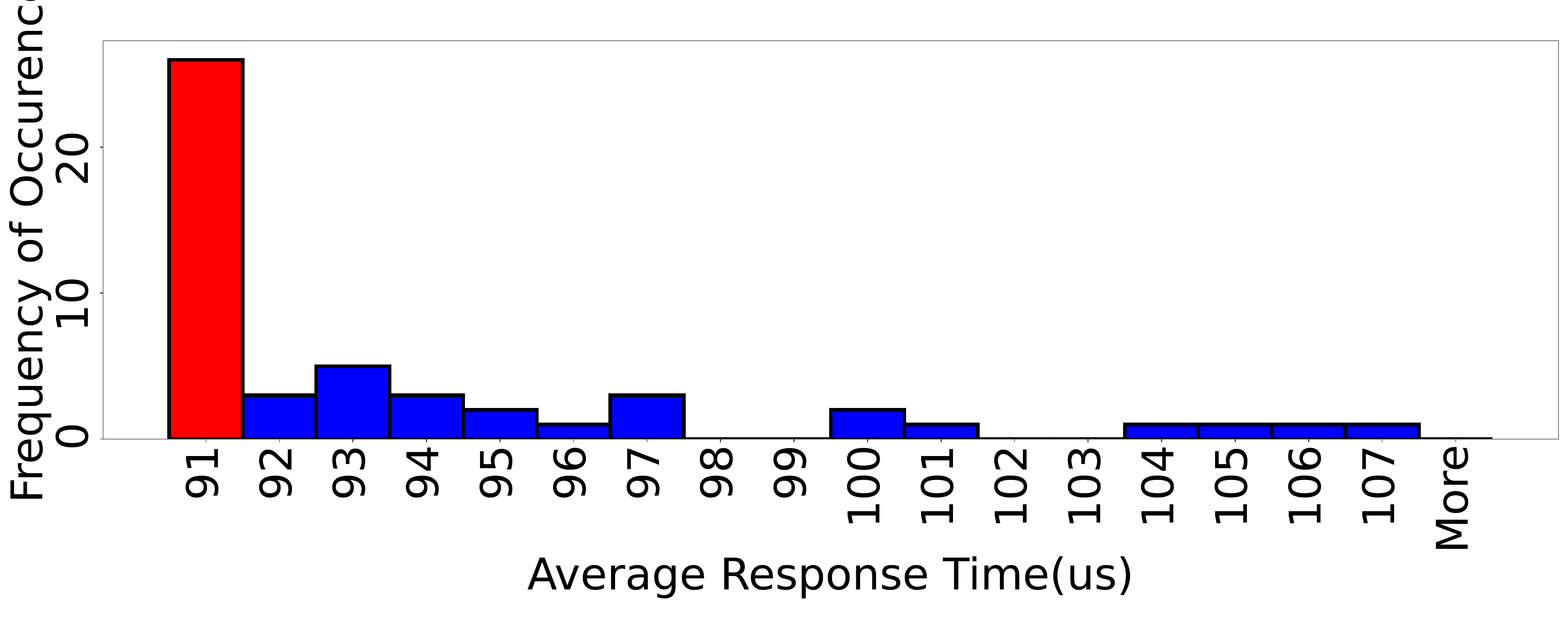}
    \caption{Frequency Chart for HP-SMToff 400K configuration. The red bar is where the median lies.}
    \label{fig:fc_hpsmtoff400k}
\end{figure}


Figure~\ref{fig:shapiro} tests the normality of the data presented earlier in Section~\ref{sec:evaluation:client-conf-impact} using the Shapiro-Wilk test. 
Data points within a single configuration correspond to varying loads (QPS), all collected from the same single server.
The red dashed line indicates the threshold below which configurations do not conform to a normal distribution. 
We analyze a total of 42 configurations (six scenarios each with seven QPS values), with each configuration comprising 50 runs. 
Approximately 50\% of these configurations adhere to a normal distribution, while the remaining 50\% do not.

The above normality test results are in line with previous work that examines data normality on a single node. 
Specifically, within the LP-SMToff scenario, all QPS configurations exhibit a normal distribution. 
Conversely, none of the QPS configurations within the HP-SMTon scenario adhere to a normal distribution. 
In the HP/LP-SMToff and HP/LP-C1Eon scenarios, approximately half of the QPS configurations follow a normal distribution, while the remaining half corresponding to the high QPS configurations do not adhere to a normal distribution.
In the HP/LP-SMToff and HP/LP-C1Eon scenarios, about half of the QPS configurations conform to a normal distribution, whereas the other half, comprising the high QPS configurations, do not.
We attribute this non-normality to queuing effects that are more pronounced for higher QPS and the reduced number of logical threads (SMToff). 
Looking into the frequency charts of these high QPS configurations, a large number of samples lies below and close to the median of the distribution, whereas a small number of samples is scattered in a larger range above the median, making the distribution skewed, as shown in Figure~\ref{fig:fc_hpsmtoff400k}. 

Based on the above normality test results, we use both parametric and non-parametric (CONFIRM) methods to calculate the number of required iterations to achieve a confidence interval with at-most 1\% error and 95\% confidence level for each configuration (as explained in Section~\ref{sec:statistics}). 



Table~\ref{tab:param_confirm_shapiro} presents the results of the two methods along with the Shapiro-Wilk test results. The highest value of iterations estimated by CONFIRM is >50 since each experiment is executed 50 times. The lowest value estimated by CONFIRM is 10, since the method assumes that smaller subsets cannot estimate non-parametric CIs reliably. The two methods do not produce exactly the same results partly due to the parametric method's ability to reliably estimate values and provide tight bounds with a fewer number of iterations, typically below 10, when the configuration adheres to a normal distribution. As a result, there are several cases where the parametric method estimates just one iteration, and the CONFIRM method requires 10 iterations. 

Nevertheless, the two methods support that different client configurations require different number of iterations to produce tight statistically confident results. For low QPS (10K - 100K), both methods agree that the LP client requires a large number of iterations to achieve statistical confidence whereas the HP client requires much less. For high QPS (300K - 500K), the HP client requires more iterations than the LP client. This behavior agrees with our empirical observations, that want the LP client to have higher standard deviation in low QPS than the HP client, and the HP client to have higher standard deviation in high QPS (see Figure~\ref{fig:smt-stdev-avg:mem-hds}).



\finding{
    The client-side hardware configuration can affect the number of iterations needed for an experiment because 
    different configurations can exhibit different levels of performance 
    \DIFadd{variability}.
    Current experimental methods 
    are effective at estimating the number of iterations required to mitigate the performance variability caused by the client.
}

\begin{table}[ht]
\caption{Number of iterations to gain statistical confidence and Shapiro-Wilk results.}
\label{tab:param_confirm_shapiro}
\resizebox{\columnwidth}{!}{%
\begin{tabular}{|c|c|c|c|c|}
\hline
\textbf{Configuration}              & \textbf{QPS} & \textbf{Parametric} & \textbf{CONFIRM} & \textbf{Shapiro-Wilk} \\ \hline
\multirow{7}{*}{\textbf{LP-SMToff}} & 10K          & 288                 & \textgreater{}50 & pass                  \\ \cline{2-5} 
                                    & 50K          & 93                  & \textgreater{}50 & pass                  \\ \cline{2-5} 
                                    & 100K         & 15                  & 37               & pass                  \\ \cline{2-5} 
                                    & 200K         & 3                   & 11               & pass                  \\ \cline{2-5} 
                                    & 300K         & 2                   & 11               & fail                  \\ \cline{2-5} 
                                    & 400K         & 5                   & 19               & fail                  \\ \cline{2-5} 
                                    & 500K         & 19                  & \textgreater{}50 & fail                  \\ \hline
\multirow{7}{*}{\textbf{LP-SMTon}}  & 10K          & 225                 & \textgreater{}50 & pass                  \\ \cline{2-5} 
                                    & 50K          & 70                  & \textgreater{}50 & pass                  \\ \cline{2-5} 
                                    & 100K         & 17                  & 34               & pass                  \\ \cline{2-5} 
                                    & 200K         & 4                   & 16               & pass                  \\ \cline{2-5} 
                                    & 300K         & 3                   & 11               & pass                  \\ \cline{2-5} 
                                    & 400K         & 4                   & 15               & pass                  \\ \cline{2-5} 
                                    & 500K         & 10                  & 36               & pass                  \\ \hline
\multirow{7}{*}{\textbf{HP-SMToff}} & 10K          & 1                   & 10               & pass                  \\ \cline{2-5} 
                                    & 50K          & 1                   & 10               & fail                  \\ \cline{2-5} 
                                    & 100K         & 1                   & 10               & pass                  \\ \cline{2-5} 
                                    & 200K         & 2                   & 11               & fail                  \\ \cline{2-5} 
                                    & 300K         & 27                  & \textgreater{}50 & fail                  \\ \cline{2-5} 
                                    & 400K         & 123                 & \textgreater{}50 & fail                  \\ \cline{2-5} 
                                    & 500K         & 203                 & \textgreater{}50 & fail                  \\ \hline
\multirow{7}{*}{\textbf{HP-SMTon}}  & 10K          & 1                   & 10               & fail                  \\ \cline{2-5} 
                                    & 50K          & 1                   & 10               & fail                  \\ \cline{2-5} 
                                    & 100K         & 1                   & 10               & fail                  \\ \cline{2-5} 
                                    & 200K         & 1                   & 10               & fail                  \\ \cline{2-5} 
                                    & 300K         & 8                   & 11               & fail                  \\ \cline{2-5} 
                                    & 400K         & 39                  & 41               & fail                  \\ \cline{2-5} 
                                    & 500K         & 77                  & 41               & fail                  \\ \hline
\multirow{7}{*}{\textbf{LP-C1Eon}}  & 10K          & 303                 & \textgreater{}50 & pass                  \\ \cline{2-5} 
                                    & 50K          & 89                  & \textgreater{}50 & pass                  \\ \cline{2-5} 
                                    & 100K         & 20                  & \textgreater{}50 & pass                  \\ \cline{2-5} 
                                    & 200K         & 3                   & 11               & pass                  \\ \cline{2-5} 
                                    & 300K         & 1                   & 10               & pass                  \\ \cline{2-5} 
                                    & 400K         & 2                   & 11               & pass                  \\ \cline{2-5} 
                                    & 500K         & 9                   & 21               & fail                  \\ \hline
\multirow{7}{*}{\textbf{HP-C1Eon}}  & 10K          & 2                   & 11               & fail                  \\ \cline{2-5} 
                                    & 50K          & 1                   & 10               & fail                  \\ \cline{2-5} 
                                    & 100K         & 1                   & 10               & pass                  \\ \cline{2-5} 
                                    & 200K         & 1                   & 10               & pass                  \\ \cline{2-5} 
                                    & 300K         & 8                   & 24               & fail                  \\ \cline{2-5} 
                                    & 400K         & 21                  & \textgreater{}50 & fail                  \\ \cline{2-5} 
                                    & 500K         & 32                  & \textgreater{}50 & pass                  \\ \hline
\end{tabular}%
}
\end{table}
\section{Configuration Recommendations}


Drawing from the taxonomy of Section~\ref{sec:clientcaused_variation} and the experimental analysis of Section~\ref{sec:evaluation}, we now discuss recommendations for how to best configure the client side in an experimental evaluation based on latency-sensitive microservices. We focus on the aspect of time-sensitivity caused by the interarrival time implementation of open-loop workload generators, as we find this can play a key role in performance 
\DIFadd{variation}.


For a time-sensitive interarrival time implementation, the client-side hardware configuration should be tuned for performance. The performance configuration mitigates the hardware timing overheads of power and energy optimizations (\ie, C-states, DVFS
), allowing the workload generator to send requests as close as possible to the time indicated by the interarrival time distribution. 
In this case 
however, it is essential to consider how accurately 
the performance configuration reflects the 
configuration within the target production cluster. If the configuration deviates from the target production configuration, then it may over- or under-estimate performance metrics, such as end-to-end time (Section~\ref{sec:evaluation:client-conf-impact}), and consequently affect any conclusions drawn, such as those related to resource provisioning.

For a time-insensitive interarrival time implementation, the choice is guided by the target environment. 
The configuration of the client should match the configuration of the target environment. When the target configuration is unknown, a space exploration could be made to evaluate a technique under several scenarios, using either homogeneous or heterogeneous client and server machine configurations. 

As far as the number of iterations required for an experiment is concerned, well established methodologies~\cite{jain:performanceanalysis:book:1991,maricq:tamingvariability:usenix:2018} should be used based on the distribution followed by the samples.

\section{Related Work}
To the best of our knowledge, this is the first study to investigate the impact of the client-side hardware configuration on the accuracy and evaluation time of an experiment. Previous works focus on quantifying variability arising from other sources, including the order of experiments, process variation, and the server-side configuration. Several studies propose techniques to mitigate variability, including increasing the number of repetitions, using confidence intervals, and developing more robust workload generators.

\subsection{Evaluating Performance using Microservices} 
In recent years, latency-critical applications have moved from a monolithic to a microservice-based software architecture to satisfy service-level objectives, availability, scalability, and regular updates \cite{lewis:microservices:2014,twitter:microservices, brigham:microservicesamazon}. In a microservice-based software architecture, an application is decomposed into several services that communicate with one another via the network through well-defined interfaces, such as gRPC and REST APIs. 
The decoupled nature of these applications leads to stricter QoS constraints per service compared to their monolithic counterparts, ranging from 250us~\cite{chou:udpm:hpca:2019,zhan:carb:cal:2016} to 500us~\cite{belay:ixos:tocs:2016,kasture:tailbench:iiswc:2016}), due to increased network communication overheads.

The transition to a microservice software paradigm has prompted the community to develop new benchmark suites, such as MicroSuite~\cite{sriraman:usuite:iiswc:2018} and DeathStar~\cite{gan:deathstar:asplos:2019}, and adopt existing services, such as Memcached~\cite{memcached}, to effectively evaluate designs targeting microservices.
Memcached, in particular, has been the focus of numerous studies, including tail-latency optimizations~\cite{Mirhosseini:qzilla:hpca:2020}, collocation~\cite{leverich:reconciling:eurosys:2014}, request scheduling and consolidation, and C-states~\cite{asyabi:peafowl:socc:2020,jawad:agilewatts:micro:2022,antoniou:agilepkgc:micro:2022}, 
due to its critical role in enhancing response times of latency-critical applications as a lightweight caching service~\cite{luo:alibaba:socc:2021}.
To simplify the experimental environment and facilitate reproducibility, deploying a single memcached server process for the experimental evaluation has been common practice among previous works~\cite{leverich:reconciling:eurosys:2014,asyabi:peafowl:socc:2020,chou:dynsleep:islped:2016,wang:smartharvest:eurosys:2021}. 


\subsection{Quantifying Performance Variability} 
Quantifying variability has been the focus of many works on datacenters, supercomputers and smartphones. Maricq \etal ~\cite{maricq:tamingvariability:usenix:2018} investigate what is the inevitable variability across nodes of the same architecture in a cluster. They conclude that variability of up-to 10\% can be attributed to the underlying hardware. Additionally, they investigate the normality of performance samples across nodes and conclude that the performance samples follow a non-parametric distribution across nodes. 
In a similar setup, Duplyakin \etal~\cite{dmitry:ordersage:usenixatc:2023} investigate the variability caused by the execution order of experiments.
The rationale is that the sequence in which experiments are conducted can alter the microarchitectural characteristics of the machine, inevitably affecting the performance outcomes of each experiment.
If executed in a specific order, this bias will impact the outcome of the experiment. Such variability can be categorized as a form of measurement bias. A measurement bias~\cite{mytkowicz:producingwrongdata:asplos:2009} is when a technique X speedups a system O by Z but the speedup is not only a result of the technique but is also a bias of the experimental setup. Several works have investigated this phenomenon in various settings, including scheduling algorithms of supercomputers~\cite{tsafrir:shaking:mascots:2007}, architectural simulations for multithreaded workloads~\cite{alameldeen:variabilitysimulators:hpca:2003}, and \DIFedit{O3 optimizations in SPEC CPU2006 workloads~\cite{mytkowicz:producingwrongdata:asplos:2009}}. Additionally to measurement bias, other  works~\cite{wright:understandingvariation:hpcmpugc:2009,alexandru:bigdatareproducible:usenix:2020} have identified the network contention as a major contribution to the variability observed by an application. Another work~\cite{jimenez:crossplatformvariation:ipdpsw:2016} has developed a methodology using the stress-ng~\cite{stress-ng} tests that can estimate the variability across machines of different architecture and as a result can reproduce with some error an experiment outcome on a different machine. 
Finally, Srinivasa \etal~\cite{prasad:smartphonesvariations:ispass:2019} manage to create a methodology that quantifies the process variation of smartphones at system level. Although different sources of performance variability have been investigated there is no mention of the configuration of the client side, even in survey papers ~\cite{hoefler:12ways:sc:2015,dmitry:ordersage:usenixatc:2023}, even though it can affect the accuracy of measurements.

\subsection{Mitigating Performance Variability} Strategies aiming to improve the experimental methodology accuracy and mitigate the performance variability have been proposed, many of which have been implemented inside a workload generator~\cite{kogias:lancet:usenix:2019,zhang:treadmill:isca:2016,leverich:reconciling:eurosys:2014,curiel:workloadsurvey:comst:2018}. For example in Lancet~\cite{kogias:lancet:usenix:2019}, the authors try to create a workload generator that accurately captures the 99th percentile latency of microservices by (i) minimizing the errors caused by excessive user interference, (ii) using state of the art hardware-based techniques, and (iii) using excessively statistical methods to accurately process the samples. Their workload implements, among others, an Anderson Darling test to check the request arrival distribution, an Augmented Dickey Fuller test to check the stationarity of samples, and a Spearman test to check whether samples are independent. Apart from workload generators, standalone tools have also been proposed, such as CONFIRM~\cite{maricq:tamingvariability:usenix:2018} and OrderSage~\cite{dmitry:ordersage:usenixatc:2023}, aiming to mitigate variability, in this case, by (i) calculating the CI for non-parametric distributions, and (ii) randomizing the order of experiments. Another set of works~\cite{wang:onlineanomalydetection:ifip:2011,zhao:cloudperformancevariabilityprediction:icpe:2021,duplyakin:datacenterconstantischange:ccgrid:2020} fall under the umbrella of variability detection (anomaly detection or changepoint detection). Other works aim to mitigate measurement bias by randomizing the experimental setup, either through slight changes in the timing of requests of the workload~\cite{tsafrir:shaking:mascots:2007} or modifications in the simulator to change cache-miss penalty~\cite{alameldeen:variabilitysimulators:hpca:2003}. \DIFadd{Finally, request batching \cite{suresh:batching:www:2019} has been proposed for eliminating network variability in Memcached.} The proposals mentioned above are complementary to our work. 

\section{Conclusions}
To the best of our knowledge, this is the first work that examines the impact of the client-side configuration on the experimental evaluation of microservices. Our evaluation reveals that under certain conditions that concern the design of the workload generator and the characteristics of a microservice, the client-side configuration can influence the accuracy  of the end-to-end measurements by up-to 150\% for the average and 200\% for the 99th percentile latency of Memcached. Motivated by the above, we provide recommendations regarding the experimental environment configuration so that any unnecessary time bias is avoided and so that the results of the experiment reflect closely the behaviour of the target environment. These results support that the client-side configuration should be considered when designing experiments. 

\section*{Acknowledgments}
The authors would like to thank the anonymous reviewers for their insightful comments on earlier versions of this manuscript. 
This project has received funding from the European Union’s Horizon 2020 research and innovation programme under the Marie Skłodowska-Curie grant agreement No 101029391.

\bibliographystyle{IEEEtranS}
\bibliography{refs}

\end{document}